%% file: multipool.tex
\documentclass[conference,a4paper]{IEEEtran}

\input{setup/preamble}  

\begin{document}
\bstctlcite{IEEEexample:BSTcontrol}

\title{Optimizing Scalable Multi-Cluster Architectures for Next-Generation Wireless Sensing and Communication}

\author{%
Samuel Riedel\textsuperscript{\ensuremath{*}}\quad
Yichao Zhang\textsuperscript{\ensuremath{*}}\quad
Marco Bertuletti\textsuperscript{\ensuremath{*}}\quad
Luca Benini\textsuperscript{\ensuremath{*\dag}}
\\
{\small
 \textsuperscript{\ensuremath{*}}IIS, ETH Z\"{u}rich\quad%
 \textsuperscript{\ensuremath{\dag}}DEI, University of Bologna%
}
\\
{\small\itshape%
 \textsuperscript{\ensuremath{*}}\{sriedel,yiczhang,mbertuletti,lbenini\}@ethz.ch%
}
}

\maketitle

\begin{abstract}
Next-generation wireless technologies (for immersive-massive communication, joint communication and sensing) demand highly parallel architectures for massive data processing. A common architectural template scales up by grouping tens to hundreds of cores into shared-memory clusters, which are then scaled out as multi-cluster manycore systems. This hierarchical design, used in GPUs and accelerators, requires a balancing act between fewer large clusters and more smaller clusters, affecting design complexity, synchronization, communication efficiency, and programmability. 
While all multi-cluster architectures must balance these trade-offs, there is limited insight into optimal cluster sizes. This paper analyzes various cluster configurations, focusing on synchronization, data movement overhead, and programmability for typical wireless sensing and communication workloads. We extend the open-source shared-memory cluster \mempool{} into a multi-cluster architecture and propose a novel double-buffering barrier that decouples processor and DMA. Our results show a single 256-core cluster can be twice as fast as 16 16-core clusters for memory-bound kernels and up to \SI{24}{\percent} faster for compute-bound kernels due to reduced synchronization and communication overheads.
\end{abstract}

\begin{IEEEkeywords}
Manycore, RISC-V, Synchronization
\end{IEEEkeywords}

\section{Introduction}\label{sec:multipool:introduction}

The IMT-2030 framework for next-generation wireless standards envisions dense integration between the \gls{gnb} and a distributed network of \gls{ai} powered intelligent actuators and sensors~\cite{itu_2030}. At least four of the six use cases of the \gls{6g} wireless standard require parallel processing of sensor data streams and \gls{ai} workloads at high data rates: \emph{immersive communication} for immersive \gls{xr} (remote multisensor telepresence and holographic communications), \emph{massive communication} (connecting massive number of devices for smart cities, transportation, logistics, environment monitoring), \emph{Artificial Intelligence and Communication} (assisted automated driving, autonomous collaboration between devices), and \emph{\gls{jcas}} (outdoor\&indoor navigation, movement tracking). \gls{jcas} represents a growing trend, aiming to exploit the wireless spectrum for sensing, requiring wider transmission bandwidth for better positioning accuracy, and multiple 64-to-256-antenna panels alternating between sensing or communication, using spatial beamforming~\cite{nokia_jsac_2023, Liu_jsac_2022}. This evolving landscape calls for programmable, massively parallel processing to support not only wireless data streams but also mixed sensing tasks, such as object detection and real-time actuator response, along with \gls{ai}-centric computation across distributed infrastructure.

Modern manycore clusters for data-parallel wireless processing typically scale by combining tightly coupled compute clusters of tens of cores into multi-cluster architectures, such as \glspl{gpu}~\cite{nvidia2020ampere, nvidia_aerial} and Kalray~\cite{dupontdedinechin2021}. However, the granularity of these clusters and the rationale for a specific choice in the number of processors in a cluster have not been studied in depth. Recent work~\cite{zhang2024} demonstrated the physical feasibility of the tightly coupled clusters up to a thousand cores. Hence, a wide design space needs to be explored (from tens to thousands of cores) to find the optimal cluster granularity.

Achieving high performance for parallel processing of multiple sensors and large-bandwidth \gls{jcas} waveforms requires balancing cluster size (scaling-up) and count (scaling-out). Tightly coupled shared-memory clusters allow efficient data-sharing, communication, and programming, but interconnects, latencies, and physically feasible routing complexity limit scaling. Beyond a certain granularity, intuitively, loosely coupling clusters becomes advantageous, utilizing \gls{dma} engines and a network-on-chip for data-transfers, despite the communication, programmability, and synchronization cost.

This paper presents a quantitative analysis of multi-cluster vs single large-cluster configurations, focusing on synchronization, data movement, and programmability of data-intensive \gls{jcas} kernels. We extend the open-source \mempool{} architecture~\cite{riedel2023mempool}, which scales a single cluster to up to 256 cores, enabling different cluster-based configurations with the same total core count. Using various double-buffered kernels for wireless sensor processing, we evaluate synchronization overhead, total core drift during computation, and communication overhead across configurations with kernels of different operational intensities.

Finally, to minimize real-time processing latency across various \gls{jcas} workloads and stages, we propose a new barrier implementation for manycore double-buffering. Instead of a naive barrier that waits for all \glspl{pe} and the \gls{dma} to complete, we decouple them: each only waits for its next buffer to be ready, not for the entire system to finish the current phase. Specifically, \glspl{pe} can proceed without waiting for all others if their next compute buffer is prepared, reducing idle time and improving performance through a simple barrier optimization. The contributions of this paper are the following:

\begin{itemize}
  \item Extending the open-source \mempool{} design into a multi-cluster design, including runtime and kernel support.
  \item A \emph{soft} double-buffering barrier, which decouples the \gls{dma} and core barriers to minimize synchronization overhead.
  \item A performance analysis on key \gls{jcas} kernels comparing different sized multi-cluster designs, balancing large cluster's intra-cluster interconnect overhead and the data splitting and synchronization overhead of multi-cluster systems.
\end{itemize}

\section{Related Work}\label{sec:multipool:related_work}

Shared-memory cluster architectures are common in modern wireless communication, large-scale sensing, and \gls{ai}-enhanced post-processing, with increasing core counts~\cite{karlrupp2022}. Many use a multi-cluster approach, but few justify the cluster size or core count ratios. Clusters are often assumed to be as large as possible, considering architectural and back-end design factors. For example, Kalray's MPPA-256 has 16 clusters of 16 cores each~\cite{dinechin2013mppa256}, and their newer MPPA DPU has 5 clusters of 16 application cores each~\cite{dinechin2022}. The CoreVA-MPSoC multiprocessor uses a similar approach, targeting software-defined radio applications, with up to 32 cores in 16-core clusters~\cite{sievers2017}. 

Architectures with programmable cores typically feature clusters with a few dozen cores, while \glspl{gpu} utilize the \gls{simt} model to expand cluster sizes. For this reason, \glspl{gpu} are widely adopted for the massively parallel artificial intelligence and next-generation wireless communication processing. NVIDIA's \gls{gpu} architecture history offers insights into cluster size and count evolution. The Fermi architecture had 16 \glspl{sm}/clusters with 32 cores each, totaling 512 cores~\cite{nvidia2010fermi}. The Kepler architecture tripled the core count by increasing cores per cluster to 192 and reducing the number of clusters~\cite{nvidia2012kepler}. Cluster size then shrank while the number of clusters grew: 128 cores for Maxwell~\cite{nvidia2014maxwell}, 64 cores for Pascal~\cite{nvidia2016pascal} and Turing~\cite{nvidia2018turing}, and 128 cores again from Ampere~\cite{nvidia2020ampere} to Ada~\cite{nvidia2023ada}. However, the reasoning behind these decisions remains unspecified.

Research on trade-offs between shared-memory systems and message passing has focused primarily on large-scale distributed systems or network topologies connecting clusters~\cite{calciu2013, lang2012, monchiero2006distributedexploration}. However, there remains a significant gap in research regarding the programming, synchronization, and communication overhead of single-chip multi-cluster designs compared to a single cluster.

\section{Architecture}\label{sec:multipool:architecture}

Evaluating multi-cluster designs requires a manycore system designed for wireless sensing and communication applications that scales to hundreds of cores. The open-source \mempool{} architecture supports up to 256 cores with \SI{1}{\mebi\byte} of L1 memory accessible within five cycles~\cite{riedel2023mempool}. Based on related work, 256 cores represent the practical limit for single-cluster designs, as no larger industrial cluster has been reported. \mempool{} uses a software-managed scratchpad memory, controlled by cores via their \gls{dma}, instead of a hardware data cache. This approach enables fine-grained, efficient data movement and allows detailed analysis without cache-related inefficiencies or noise.

\begin{figure}[tb]
  \centering
  \includegraphics[width=\columnwidth]{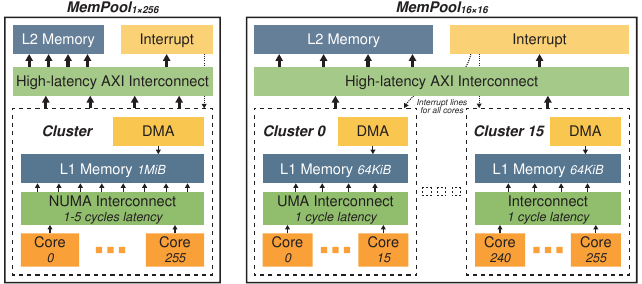}
  \caption{Architecture of \mempool{} (or \multipool{1}{256}) (left) and the multi-cluster extension illustrated on \multipool{16}{16} (right). Both systems have the same L2 bandwidth and access to a global interrupt controller with dedicated lines to all clusters.}\label{fig:multipool:architecture}
  \vspace*{-\baselineskip}
\end{figure}

We extend \mempool{} to support cluster replication while maintaining the same system interconnect, enabling various 256-core configurations as shown in \cref{fig:multipool:architecture}: from a single 256-core cluster (\multipool{1}{256}) to multiple smaller clusters, e.g., 16 clusters of 16 cores each (\multipool{16}{16}). The total number of cores and L1 memory remains constant, but L1 memory is split among clusters. Small clusters feature single-cycle L1 interconnect latency, while larger clusters require a \gls{numa} scheme with up to 5-cycle latency. Each cluster has a private \gls{dma} engine for L1–L2 transfers, and the system interconnect links all clusters to a constant-bandwidth L2 memory (on-chip \acrshort{sram} or off-chip \acrshort{dram}).

Clusters synchronize using \glspl{amo} on private L1 or shared L2 memory. Cores can also enter wait-for-interrupt states and trigger interrupts via a system-wide controller, enabling intra- and inter-cluster barriers.

\section{Barriers in Double-Buffering}\label{sec:multipool:barriers}

\begin{figure}[thb]
  \centering
  \includegraphics[width=\columnwidth]{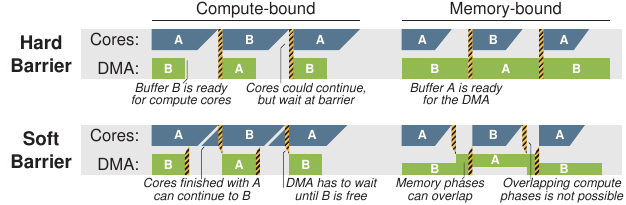}
  \caption{Timeline of double-buffered executions for compute-bound (left) and memory-bound (right) executions with \emph{hard} barriers (top) or \emph{soft} barriers (bottom). Blue boxes represent compute phases, with slanted boxes indicating varying start and finish times for cores. Green boxes denote DMA transfers, and bee-colored lines show phase barriers.}\label{fig:multipool:barrier_diagram}
  \vspace*{-\baselineskip}
\end{figure}

To evaluate multi-cluster synchronization and data-movement overhead, kernels are distributed across clusters and evaluated with their transfer phases, often using double-buffered execution to overlap memory transfers with processing (see \cref{fig:multipool:barrier_diagram}). Two buffers, \emph{A} and \emph{B}, are used: the \emph{transfer} buffer is refilled by the \gls{dma}, while the \emph{compute} buffer is processed by the cores. Once done, the buffers switch roles, and the \gls{dma} refills the old compute buffer while cores process the new one.

Performance is limited by either processing or transfer time, depending on kernel arithmetic intensity, system performance, and bandwidth, resulting in compute-bound or memory-bound execution. In compute-bound cases (\cref{fig:multipool:barrier_diagram} left), memory transfers finish first, causing the \gls{dma} to wait. In memory-bound cases, cores finish first and wait for the \gls{dma}.

\subsection{Soft Barriers}\label{subsec:multipool:soft_barriers}

The double-buffering scheme requires synchronization, typically through a barrier, to ensure two dependencies: the \gls{dma} completes writing results from L1 to L2 and loads the next phase buffer from L2 to L1, while all cores finish computation for result transfer~\cite{sancho2008}. This \emph{hard} barrier prevents any actor from advancing to the next phase until all processes of the current phase are completed, ensuring the system is synchronized.

The synchronization guarantee can be relaxed by \emph{softening} the barrier, as shown in \cref{fig:multipool:barrier_diagram}. In both compute-bound and memory-bound executions, either \glspl{dma} or cores idle at phase ends, with buffers ready for the next phase. For compute-bound kernels, once the \gls{dma} finishes, cores can proceed even if not all are done, reclaiming lost performance due to resource contention or unbalanced parallelization. A \emph{soft} barrier enables overlapping compute phases, reducing overall runtime as shown in the bottom left of \cref{fig:multipool:barrier_diagram}. In memory-bound kernels, the \gls{dma} waits for computation to complete. While overlapping compute phases is not possible here, the previous compute buffer can already be processed by the \gls{dma}. In scenarios like decoupled in/out channels (e.g., with the \gls{axi} protocol) or multiple \glspl{dma}, the \emph{soft} barrier can still improve performance, even in memory-bound kernels.

While the \gls{dma} cannot leverage overlapping transfer phases in \mempool{}, compute-bound kernels can benefit from \emph{soft} barriers due to high core count and execution time variance. The soft barrier is implemented as follows: at the end of computation, cores atomically increment a barrier counter and check the core count. They then wait for \gls{dma} completion. The first core polls the \gls{dma} termination signal and sends an interrupt to the other cores, signaling them to proceed. The last core resets the barrier and initiates the next transfer, moving the \gls{dma} to the next phase. All actors then transition to the next phase.

An additional communication step prevents any core from advancing two stages ahead. After completing the previous phase, the last core sends an interrupt to signal that the \gls{dma} is programmed and the barrier is reset. Cores check this signal before entering the next barrier to ensure safe synchronization.

\section{Evaluation Setup}\label{sec:multipool:evaluation_setup}

\begin{figure}[t]
  \centering
  \includegraphics[width=\linewidth]{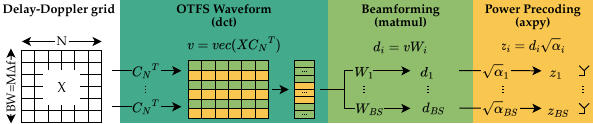}
  \caption{DCT-based waveform generation for \gls{otfs} \gls{jcas}.}\label{fig:multipool:application}
  \vspace*{-\baselineskip}
\end{figure}

To benchmark the clusters, we follow the \gls{otfs} proposal for \gls{jcas} depicted in \cref{fig:multipool:application} and benchmark the corresponding kernels~\cite{Kalpage2024, Li2022}. We run the following benchmarks using cycle-accurate \acrlong{rtl} simulations, generating deterministic execution traces for both cores and \glspl{dma}. These traces enable precise timelines, instruction breakdowns, and performance analysis. Our benchmarks use double-buffered 32-bit integer kernels common in the wireless sensing and computation domain, implemented with \emph{hard} and \emph{soft} barriers, parallelized across all cores in each cluster. The total memory across clusters is constant, so problem sizes remain fixed for all kernels, except for matrix multiplication.

\setlist[description]{font=\normalfont}

\textbf{\dct{}:}
As shown in \cref{fig:multipool:application}, the \gls{dct} is a key kernel in \gls{otfs} modulation for \gls{jcas} as described in~\cite{Kalpage2024}. Each core processes its own 64 points in parallel. The input buffer can be overwritten with the output buffer, allowing a doppler frame of \by{96}{1024} per iteration.

\textbf{\matmul{}:}
Beamforming (matrix–matrix multiplication) is essential for directing signal energy to enhance transmission efficiency and positioning precision. To fit 6 matrix buffers into L1 memory, \multipool{1}{256} operates on \by{192}{192} matrices, and smaller clusters on \by{128}{128}, \by{96}{96}, \by{64}{64}, and \by{48}{48} matrices. The tiling impacts arithmetic intensity. In \multipool{16}{16}, 16 clusters compute a \by{192}{192} output but perform only 48 inner iterations---requiring 4$\times$ more iterations than \multipool{1}{256} to complete the \by{192}{192} matrix.

\textbf{\axpy{}:}
This kernel is defined as $\vec{y} = \alpha \cdot \vec{x} + \vec{y}$ and serves the per-antenna power precoding in \gls{jcas}. We operate on vectors of \num{49152} elements per iteration, distributed across clusters, with each core processing a chunk of the vectors.

We also benchmark two key kernels that, together with \matmul{}, are crucial in \gls{ai}-enhanced sensor processing:

\textbf{\conv{}:}
A 2D convolution commonly used for deep-learning-based channel-estimation, and object detection/classification for example in autonomous vehicles. We implement a $3 \times 3$ kernel, parallelized in an output-stationary manner, where each core computes an output tile independently. The problem size is \by{48}{1024} per iteration, divided over all clusters.

\textbf{\dotp{}:}
The dot product computes the scalar product of two vectors parallelized like \axpy{} with a problem size of \num{49152} elements. Each core sums into a private reduction variable. In the end, the sums are hierarchically reduced using \glspl{amo}: cores reduce within their cluster, and one core of each cluster adds the result to the global sum.

\begin{figure*}[tbh]
  \begin{minipage}{\columnwidth}
    \centering
    \subfloat{%
      \includegraphics[width=0.95\columnwidth]{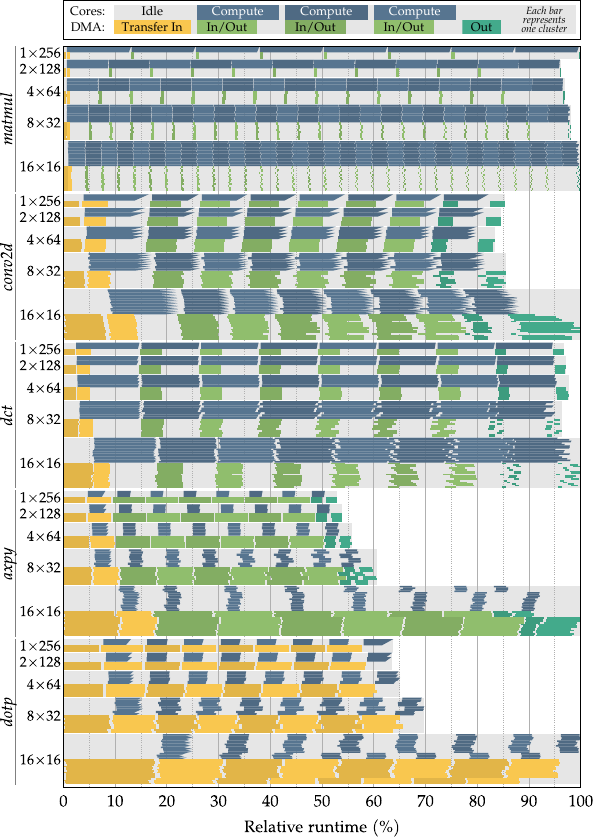}}
  \end{minipage}\hfill%
  \begin{minipage}{\columnwidth}
    \centering
    \subfloat{%
      \includegraphics[width=0.95\columnwidth]{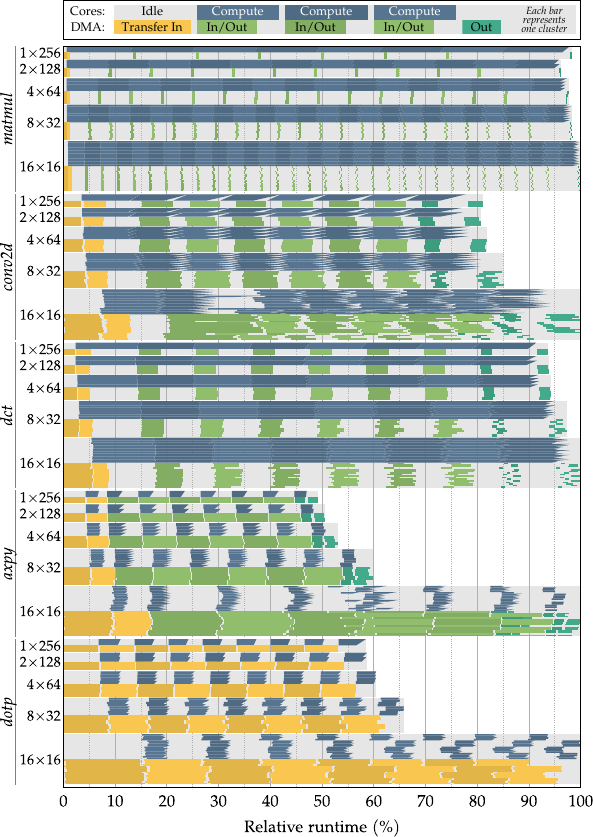}}
  \end{minipage}%
  \caption{Timeline of the kernels with \textbf{hard} barriers (left) and \textbf{soft} barriers (right) on different multi-cluster configurations. The blue lines represent the clusters' active compute phases. The yellow and green lines represent the clusters' \gls{dma} activity.}%
    \label{fig:multipool:timeline}
    \vspace*{-\baselineskip}
\end{figure*}

\section{Results}\label{sec:multipool:results}

\subsection{Timelines}

We run all double-buffered \gls{jcas} kernels with eight iterations (more for \matmul{} to account for small-cluster tiling) across various multi-cluster configurations and plot their normalized execution timelines in \cref{fig:multipool:timeline}, similar to \cref{fig:multipool:barrier_diagram}. Each configuration's bar shows two lines per cluster: the top blue line depicts compute phases as trapezoids (start of first core, all cores executing, end of last core), and the bottom line shows \gls{dma} transfers in yellow for in, light green for in/out, dark green for out phases. Odd and even phases use different shades for clarity, and the gray background indicates total execution time.

Given the information density, we analyze it step by step, starting with key messages and then detailed aspects using dedicated plots. For hard barriers (left), configurations with fewer, larger clusters consistently outperform smaller ones across all kernels. Compute-bound kernels like \matmul{}, \conv{}, and \dct{} see \SIrange{4}{15}{\percent} gains with larger clusters, though improvements saturate or worsen for the largest cluster. Memory-bound kernels like \axpy{} and \dotp{} show up to \SI{89}{\percent} gains due to transfer phase differences and global barrier effects (see \cref{subsec:multipool:results:drift}).

Comparing the hard and soft barrier timelines in \cref{fig:multipool:timeline} reveals that soft barriers notably benefit compute-bound kernels by eliminating gaps between compute phases and enabling overlap when the \gls{dma} finishes early. This leads to better performance, particularly for large clusters, as seen in \conv{}, where \multipool{2}{128} achieves a \SI{24}{\percent} speedup. Among soft barrier configurations, larger clusters again perform better, with \multipool{16}{16} consistently the slowest. Compute-bound kernels like \matmul{} and \dct{} show \SIrange{4}{7}{\percent} speedups as cluster counts decrease, with \multipool{4}{64}, \multipool{2}{128}, and \multipool{1}{256} performing similarly---although for \matmul{} and \conv{}, performance declines slightly at \multipool{1}{256}.

The memory-bound kernels show an even larger preference for a large cluster. The biggest step comes from going from \multipool{16}{16} to \multipool{8}{32}. For \axpy{} and \dotp{}, \multipool{1}{256} performs best with a speedup of 2$\times$ and 1.7$\times$, respectively.

\subsection{Steady Phase}

Although the timelines give a complete view, the steady middle phases are the most relevant. Our choice of eight phases provides a good overview, but real wireless applications mostly have more phases due to larger problem sizes, where the middle phases (simultaneous data transfer and computation) dominate execution and are therefore the focus of this section.

\subsubsection{Barrier Types}

\begin{figure}[tb]
  \centering
  \includegraphics[width=\columnwidth]{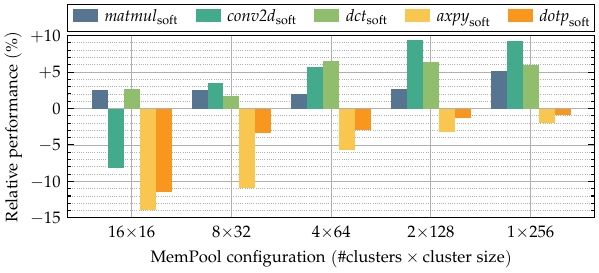}
  \caption{Relative performance of different barrier implementations for all \mempool{} configurations, normalized to the \emph{hard} barrier performance of the same configuration and kernel.}\label{fig:multipool:relpoint_performance}
  \vspace*{-1.28\baselineskip}
\end{figure}

\Cref{fig:multipool:relpoint_performance} compares soft and hard barriers, revealing two main trends. First, soft barriers outperform hard barriers by up to \SI{9}{\percent} in compute-bound kernels but underperform in memory-bound ones. In \mempool{}, merging compute phases boosts throughput, but in memory-bound cases, the increased wake-up latency after the barrier and contention on the AXI bus degrade performance. With soft barriers, cores start polling the \gls{dma} as soon as the first core finishes, competing with ongoing data transfers on the AXI bus---unlike hard barriers, which only poll after all cores complete. In compute-bound kernels, this contention is avoided as no transfers are active during polling.

The second key insight is that performance gains from soft barriers grow with cluster size. Larger clusters experience more drift and idling at hard barriers, making them benefit more from overlapping compute phases. On the other hand, for more clusters, since clusters operate independently, even hard barriers allow faster clusters to proceed without waiting for the system's slowest core, resembling soft barrier behavior.

\subsubsection{Cluster Size}

\begin{figure}[tb]
  \centering
  \includegraphics[width=\columnwidth]{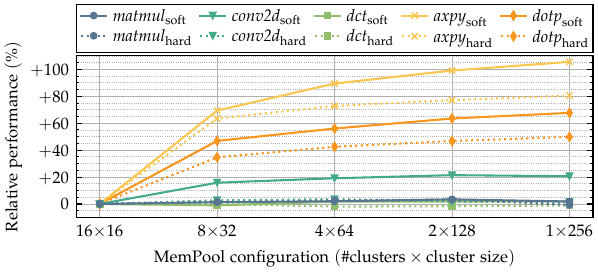}
  \caption{Relative performance comparison of different \mempool{} configurations. The data points are normalized to the \multipool{16}{16} configuration of the respective barrier implementation.}\label{fig:multipool:performance}
  \vspace*{-\baselineskip}
\end{figure}

\Cref{fig:multipool:performance} compares steady-phase performance relative to the \multipool{16}{16} configuration. For the memory-bound \axpy{} and \dotp{} kernels, larger clusters significantly boost performance by utilizing L2 bandwidth more efficiently through large, latency-tolerant \gls{dma} transfers. Soft barriers show even larger gains due to the slower baseline of \multipool{16}{16}. In \conv{}, soft barriers improve performance by about \SI{20}{\percent}, while hard barriers show little change due to higher synchronization overhead. Similarly, the compute-bound \dct{} and \matmul{} kernels gain roughly \SI{4}{\percent} with larger clusters.

\begin{figure}[tb]
  \begin{minipage}{\columnwidth}
    \centering
    \subfloat{%
      \includegraphics[width=\columnwidth]{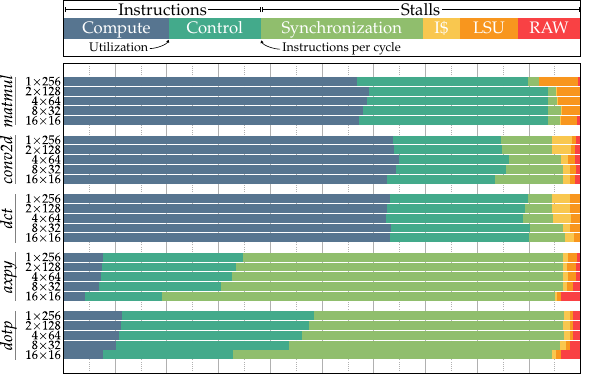}}
  \end{minipage}\vfill%
  \vspace*{0.1\baselineskip}
  \begin{minipage}{\columnwidth}
    \centering
    \subfloat{%
      \includegraphics[width=\columnwidth]{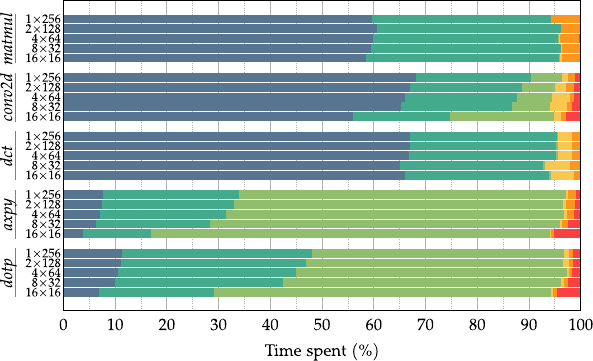}}
  \end{minipage}%
  \caption{Breakdown of cores' activity with \textbf{hard} (top) and \textbf{soft} barriers (bottom). The first two bars show time spent on compute (e.g., additions, multiplications) and control (e.g., loads, branches) instructions. Remaining bars capture idle phases: synchronization (barrier sleep), I\$ stalls, LSU stalls (interconnect congestion), and RAW stalls (read-after-write hazards). Metrics include compute unit utilization, IPC, and synchronization overhead.}%
    \label{fig:multipool:breakdown}
    \vspace*{-\baselineskip}
\end{figure}

\subsubsection{Breakdown}

To understand performance differences, \cref{fig:multipool:breakdown} analyzes core activity with hard and soft barriers. Compute-bound kernels show high compute utilization, while memory-bound kernels are dominated by data-transfer stalls. In \matmul{} with hard barriers, larger clusters experience more stalls from higher latency and lower bandwidth, but benefit from reduced synchronization time. These opposing effects lead to slightly better instruction rates for smaller clusters. However, smaller clusters incur more control overhead due to smaller problem sizes and frequent \gls{dma} calls, making medium-sized clusters the best performers. For the largest cluster, stalls dominate, reducing compute utilization---a trend also seen in other compute-bound kernels.

Comparing hard and soft barriers shows that soft barriers reduce synchronization overhead and improve compute utilization, favoring larger clusters. This shifts the \multipool{1}{256} configuration closer to or even above medium-sized clusters. Notably, in \conv{}, the fastest cores become partly memory-bound while slower ones remain compute-bound, leaving residual synchronization time, particularly in smaller clusters.

\subsection{Drift}%
\label{subsec:multipool:results:drift}

\begin{figure}[tbh]
  \vspace*{0.5\baselineskip}
  \centering
  \includegraphics[width=\columnwidth]{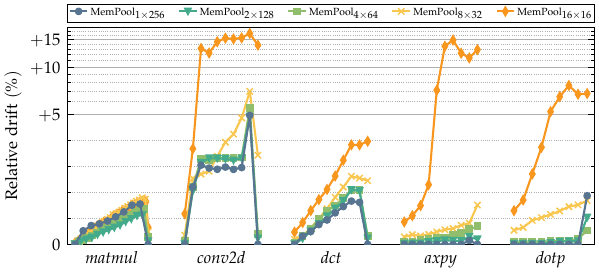}
  \caption{The time drift of each kernel with \textbf{soft} barriers is shown relative to the total pipeline execution time. The y-axis is linear up to \SI{5}{\percent}, then logarithmic to show all nuances.}%
  \label{fig:multipool:soft_drift}
  \vspace*{-\baselineskip}
\end{figure}

Most execution time is spent in steady phases, with a final synchronization phase requiring coordination across all cores, which depends on drift accumulated between clusters and cores during previous phases. \cref{fig:multipool:soft_drift} shows this drift as a percentage of total runtime, with points representing cumulative drift across all phases, from the first \gls{dma} transfer to the final \gls{dma} out phase, except for \dotp{}, which lacks the final \gls{dma} phase.

Large clusters show small, constant drifts due to internal synchronization, while more clusters experience larger drifts. Compute-bound kernels maintain small drifts, as inter-core drift determines phase ends, while memory-bound kernels have drifts between \SIrange{8}{16}{\percent}. The \multipool{1}{256} memory-bound kernels show no drift, as a single \gls{dma} phase dictates phase end. For \dotp{}, the final phase shows significant drift in the \multipool{1}{256} configuration due to a non-hierarchical reduction, while multi-cluster configurations reduce this through hierarchical reduction. The \multipool{16}{16} configuration has minimal reduction overhead, as its 16-core clusters reduce quickly, and global reduction is hidden by the clusters' drift.

\section{Cluster Scaling Limits}\label{sec:multipool:scalability_limits}

While larger clusters improve communication and synchronization overheads, their scalability is ultimately limited. In \mempool{}, gains saturate at a few hundred cores as physical design challenges emerge: maintaining frequency with increasing cluster size requires higher L1 memory latency, lower bandwidth, or both. Additionally, interconnects consume more area and power, reducing area and energy efficiency; in 1000+ core clusters, a significant portion of chip resources is spent on the hierarchical interconnect rather than on compute or memory~\cite{zhang2024}.

Higher latency and lower bandwidth exacerbate \gls{numa} issues, requiring careful optimization of memory access patterns and latency hiding~\cite{bertuletti2023pusch}. Synchronization also becomes costlier with more cores, necessitating hierarchical mechanisms~\cite{bertuletti2023barrier}. Despite software optimizations, larger clusters still suffer more stalls due to these limitations, as evident in \matmul{}'s performance.

Our analysis shows that clusters with several hundred cores are optimal in multi-cluster systems. Specifically, compute-bound kernels perform best at 128 cores per cluster, while memory-bound kernels benefit from 256-core clusters. Although scaling challenges persist, implementations up to 1024 cores are physically viable~\cite{zhang2024}, suggesting that the optimal cluster size may shift into the thousands with future technology scaling.

\section{Conclusion}\label{sec:multipool:conclusion}

This paper explores multi-cluster trade-offs of wireless manycore architectures for \gls{jcas} by extending the open-source \mempool{} architecture to a multi-cluster system. We compared a single 256-core cluster with various multi-cluster configurations, keeping the total core count constant. A single large cluster is up to 2$\times$ faster than 16 smaller ones for memory-bound wireless sensing kernels due to better bandwidth utilization. Compute-bound \gls{jcas} kernels also favor larger clusters, achieving up to \SI{24}{\percent} higher performance. The largest performance gains occur when consolidating many small clusters into fewer medium-sized ones, with diminishing returns as cluster size grows. At larger sizes, increased latency, reduced bandwidth, and physical design complexity counterbalance benefits like improved utilization and lower synchronization and communication overheads. Core activity and synchronization drift analysis show that smaller clusters experience more drift and higher synchronization costs. To mitigate this, we introduced a double-buffering barrier scheme, achieving \SIrange{5}{9}{\percent} speedups.

\section*{Acknowledgment}
This work has received funding from the Swiss State Secretariat for Education, Research, and Innovation (SERI) under the SwissChips initiative.

\Urlmuskip=0mu plus 1mu\relax
\def\UrlBreaks{\do\/\do-}
\bibliographystyle{IEEEtran}
\bibliography{bib/bstcontrol, bib/multipool_fixed}

\end{document}

%% file: setup/preamble.tex
\usepackage{cite}
\usepackage{amsmath,amssymb,amsfonts}
\usepackage{algorithmic}
\usepackage{listings}
\usepackage{graphicx}
\usepackage{subfig}
\usepackage{enumitem}
\usepackage{textcomp}
\usepackage[dvipsnames]{xcolor}
\usepackage[a4paper, total={184mm,239mm}]{geometry}
\def\BibTeX{{\rm B\kern-.05em{\sc i\kern-.025em b}\kern-.08em
    T\kern-.1667em\lower.7ex\hbox{E}\kern-.125emX}}
\usepackage{glossaries}
\usepackage[hidelinks]{hyperref}
\usepackage[capitalise,nameinlink]{cleveref}
\usepackage[stretch=40,shrink=40,step=4]{microtype}
\usepackage[free-standing-units,per-mode=repeated-symbol,mode=text,detect-weight=true,detect-family=true]{siunitx}
\usepackage[shortcuts]{extdash}

\input{setup/acronyms}

\definecolor{grey0}{HTML}{1B2B34}
\definecolor{grey1}{HTML}{343D46}
\definecolor{grey2}{HTML}{4F5B66}
\definecolor{grey3}{HTML}{65737E}
\definecolor{grey4}{HTML}{A7ADBA}
\definecolor{grey5}{HTML}{8CABA8}
\definecolor{grey6}{HTML}{BECBD2}
\definecolor{grey7}{HTML}{E9EEF0}
\definecolor{red}{HTML}{A8322D}
\definecolor{orange}{HTML}{D97531}
\definecolor{yellow}{HTML}{F8BD3F}
\definecolor{green}{HTML}{92BA51}
\definecolor{teal}{HTML}{108A7C}
\definecolor{blue}{HTML}{2A4857}
\definecolor{purple}{HTML}{875A78}
\definecolor{brown}{HTML}{805A4D}

\definecolor{color1}{HTML}{f94144}
\definecolor{color2}{HTML}{f9c74f}
\definecolor{color3}{HTML}{90be6d}
\definecolor{color4}{HTML}{43aa8b}
\definecolor{color5}{HTML}{577590}

\definecolor{PulpGreen}{HTML}{168638}
\definecolor{PulpBlue}{HTML}{1269b0}
\definecolor{PulpRed}{HTML}{a8322c}
\definecolor{PulpYellow}{HTML}{f2c100}

\PackageWarning{Citation missing:}{#1!}

\PackageWarning{TODO:}{#1!}

\newcommand\mempool{Mem\-Pool}

\newcommand\by[2]{#1$\times$#2}

\newcommand\matmul{\emph{matmul}}
\newcommand\conv{\emph{2dconv}}
\newcommand\dct{\emph{dct}}
\newcommand\dotp{\emph{dotp}}
\newcommand\axpy{\emph{axpy}}

\newcommand{\multipool}[2]{\mempool{}\ensuremath{_{\text{#1}\times\text{#2}}}}

\DeclareSIUnit\core{core}
\DeclareSIUnit\tile{tile}
\DeclareSIUnit\request{req}
\DeclareSIUnit\cycle{cycle}
\DeclareSIUnit\erlang{E}
\DeclareSIUnit\flop{FLOP}
\DeclareSIUnit\flops{FLOPS}
\DeclareSIUnit\gate{GE}
\DeclareSIUnit\ge{GE}
\DeclareSIUnit\op{OP}
\DeclareSIUnit\ops{OPS}
\DeclareSIUnit\bps{bps}
\DeclareSIUnit\Bps{Bps}
\DeclareSIUnit\ipc{IPC}
\DeclareSIUnit\ips{IPS}
\DeclareSIUnit\bits{bits}
\DeclareSIUnit\pixel{pixel}
\DeclareSIUnit{\nounit}{\relax}
\DeclareSIUnit[quantity-product=]\percent{\%}

%% file: setup/acronyms.tex
\newacronym[longplural={general-purpose \acrlongpl{gpu}}]{gpgpu}{GPGPU}{general-purpose \gls{gpu}}
\newacronym[longplural={networks-on-chip}]{noc}{NoC}{network-on-chip}
\newacronym[longplural={scratchpad memories}]{spm}{SPM}{scratchpad memory}
\newacronym{abi}{ABI}{application binary interface}
\newacronym{ace}{ACE}{AXI Coherent Extensions}
\newacronym{ai}{AI}{artificial intelligence}
\newacronym{alu}{ALU}{arithmetic logic unit}
\newacronym{amba}{AMBA}{Advanced Microcontroller Bus Architecture}
\newacronym{amo}{AMO}{atomic memory operation}
\newacronym{aot}{AOT}{ahead-of-time}
\newacronym{apb}{APB}{Advanced Peripheral Bus}
\newacronym{api}{API}{application programming interface}
\newacronym{asic}{ASIC}{application-specific integrated circuit}
\newacronym{axi}{AXI}{Advanced eXtensible Interface}
\newacronym{bfs}{BFS}{breadth-first search}
\newacronym{blas}{BLAS}{Basic Linear Algebra Subprograms}
\newacronym{cas}{CAS}{compare-and-swap}
\newacronym{cgra}{CGRA}{coarse-grained reconfigurable architectures}
\newacronym{cmos}{CMOS}{complementary metal-oxide-semiconductor}
\newacronym{cnn}{CNN}{convolutional neural network}
\newacronym{cpu}{CPU}{central processing unit}
\newacronym{csr}{CSR}{control and status register}
\newacronym{cs}{CS}{critical section}
\newacronym{dbt}{DBT}{dynamic binary translation}
\newacronym{dct}{DCT}{discrete cosine transform}
\newacronym{dlp}{DLP}{data level parallelism}
\newacronym{dma}{DMA}{direct memory access}
\newacronym{dram}{DRAM}{dynamic random-access memory}
\newacronym{dsl}{DSL}{domain-specific language}
\newacronym{dsp}{DSP}{digital signal processing}
\newacronym{elf}{ELF}{Executable and Linkable Format}
\newacronym{fdsoi}{FD-SOI}{fully depleted silicon-on-insulator}
\newacronym{fifo}{FIFO}{first in, first out}
\newacronym{fpga}{FPGA}{field-programmable gate array}
\newacronym{fpu}{FPU}{floating-point unit}
\newacronym{fp}{FP}{floating-point}
\newacronym{fsm}{FSM}{finite-state machine}
\newacronym{gpu}{GPU}{graphics processing unit}
\newacronym{hart}{hart}{hardware thread}
\newacronym{hbm}{HBM}{High Bandwidth Memory}
\newacronym{hdl}{HDL}{hardware description language}
\newacronym{hero}{HERO}{Heterogeneous Embedded Research Platform}
\newacronym{hpc}{HPC}{high-performance computing}
\newacronym{ilp}{ILP}{instruction level parallelism}
\newacronym{iot}{IoT}{Internet of Things}
\newacronym{ipc}{IPC}{instructions per cycle}
\newacronym{ipu}{IPU}{image processing unit}
\newacronym{ir}{IR}{intermediate representation}
\newacronym{isa}{ISA}{instruction set architecture}
\newacronym{issr}{ISSR}{indirection stream semantic register}
\newacronym{jit}{JIT}{just-in-time}
\newacronym{llc}{LLC}{last-level cache}
\newacronym{lrsc}{LRSC}{load-reserved/store-conditional}
\newacronym{lr}{LR}{load-reserved}
\newacronym{lsu}{LSU}{load-store unit}
\newacronym{mac}{MAC}{multiply–accumulate}
\newacronym{mcs}{MCS}{Mellor-Crummey, Scott}
\newacronym{mimd}{MIMD}{multiple instruction, multiple data}
\newacronym{mmu}{MMU}{memory management unit}
\newacronym{mpmc}{MPMC}{multi producer, multi consumer}
\newacronym{mwait}{MWait}{Memory Wait}
\newacronym{nuca}{NUCA}{non-uniform cache architecture}
\newacronym{numa}{NUMA}{non-uniform memory access}
\newacronym{os}{OS}{operating system}
\newacronym{pc}{PC}{program counter}
\newacronym{pe}{PE}{processing element}
\newacronym{pl}{PL}{programmable logic}
\newacronym{pmca}{PMCA}{programmable manycore accelerator}
\newacronym{ppa}{PPA}{power, performance and area}
\newacronym{psl}{PSL}{Power Service Layer}
\newacronym{pulp}{PULP}{Parallel Ultra Low Power}
\newacronym{qlr}{QLR}{queue-linked register}
\newacronym{qnode}{Qnode}{queue node}
\newacronym{raw}{RAW}{read-after-write}
\newacronym{rmw}{RMW}{read–modify–write}
\newacronym{rob}{ROB}{reorder buffer}
\newacronym{rom}{ROM}{read-only memory}
\newacronym{ro}{RO}{read-only}
\newacronym{rsf}{RSF}{Request-Store-Forward}
\newacronym{rtl}{RTL}{register-transfer level}
\newacronym{rvwmo}{RVWMO}{RISC-V Weak Memory Ordering}
\newacronym{sbt}{SBT}{static binary translation}
\newacronym{scm}{SCM}{standard cell memory}
\newacronym{sc}{SC}{store-conditional}
\newacronym{sdf}{SDF}{Standard Delay Format}
\newacronym{simd}{SIMD}{single instruction, multiple data}
\newacronym{simt}{SIMT}{single instruction, multiple thread}
\newacronym{sm}{SM}{streaming multiprocessor}
\newacronym{soa}{SoA}{state-of-the-art}
\newacronym{soc}{SoC}{system-on-chip}
\newacronym{sram}{SRAM}{static random-access memory}
\newacronym{ssa}{SSA}{static single assignment}
\newacronym{ssr}{SSR}{stream semantic register}
\newacronym{tas}{TAS}{test-and-set}
\newacronym{tcdm}{TCDM}{tightly-coupled data memory}
\newacronym{tlp}{TLP}{thread level parallelism}
\newacronym{uma}{UMA}{uniform memory access}
\newacronym{vliw}{VLIW}{very long instruction word}
\newacronym{vnb}{VNB}{von Neumann bottleneck}
\newacronym{vpu}{VPU}{vector processing unit}
\newacronym{war}{WAR}{write-after-read}
\newacronym{waw}{WAW}{write-after-write}
\newacronym{xr}{XR}{extended reality}
\newacronym{gnb}{gNB}{Next-Generation Node B}
\newacronym{6g}{6G}{6th generation}
\newacronym{jcas}{JCAS}{joint communication and sensing}
\newacronym{otfs}{OTFS}{Orthogonal Time Frequency Space}

%% file: multipool.bbl
\begin{thebibliography}{10}
\providecommand{\url}[1]{#1}
\csname url@samestyle\endcsname
\providecommand{\newblock}{\relax}
\providecommand{\bibinfo}[2]{#2}
\providecommand{\BIBentrySTDinterwordspacing}{\spaceskip=0pt\relax}
\providecommand{\BIBentryALTinterwordstretchfactor}{4}
\providecommand{\BIBentryALTinterwordspacing}{\spaceskip=\fontdimen2\font plus
\BIBentryALTinterwordstretchfactor\fontdimen3\font minus \fontdimen4\font\relax}
\providecommand{\BIBforeignlanguage}[2]{{%
\expandafter\ifx\csname l@#1\endcsname\relax
\typeout{** WARNING: IEEEtran.bst: No hyphenation pattern has been}%
\typeout{** loaded for the language `#1'. Using the pattern for}%
\typeout{** the default language instead.}%
\else
\language=\csname l@#1\endcsname
\fi
#2}}
\providecommand{\BIBdecl}{\relax}
\BIBdecl

\bibitem{itu_2030}
{ITU}, ``Recommendation {ITU-R M.2160}: Framework and overall objectives of the future development of {IMT} for 2030 and beyond,'' \url{https://www.itu.int/rec/R-REC-M.2160-0-202311-I/en}, 2023.

\bibitem{nokia_jsac_2023}
T.~Wild, V.~Braun, and H.~Viswanathan, ``Joint design of communication and sensing for beyond {5G} and {6G} systems,'' \url{https://www.nokia.com/bell-labs/research/6g-networks/6g-technologies/network-as-a-sensor/}, 2023.

\bibitem{Liu_jsac_2022}
F.~Liu \emph{et~al.}, ``Integrated sensing and communications: Toward dual-functional wireless networks for 6g and beyond,'' \emph{IEEE Journal on Selected Areas in Communications}, vol.~40, no.~6, pp. 1728--1767, 2022.

\bibitem{nvidia2020ampere}
{NVIDIA Corporation}, ``{NVIDIA} {Ampere} {GA102} {GPU} architecture,'' \url{https://www.nvidia.com/content/PDF/nvidia-ampere-ga-102-gpu-architecture-whitepaper-v2.pdf}, NVIDIA Corp., Tech. Rep., 2020.

\bibitem{nvidia_aerial}
A.~Kelkar and C.~Dick, ``Nvidia aerial gpu hosted ai-on-5g,'' in \emph{2021 IEEE 4th 5G World Forum (5GWF)}, 2021, pp. 64--69.

\bibitem{dupontdedinechin2021}
B.~{Dupont de Dinechin}, ``A qualitative approach to many‐core architecture,'' in \emph{dupo}, L.~Andrade and F.~Rousseau, Eds.\hskip 1em plus 0.5em minus 0.4em\relax Hoboken, New Jersey, USA: Wiley, Apr. 2021, ch.~2, pp. 27--51.

\bibitem{zhang2024}
Y.~Zhang, M.~Bertuletti, S.~Riedel, M.~Cavalcante, A.~Vanelli-Coralli, and L.~Benini, ``{TeraPool-SDR}: An {1.89TOPS} 1024 {RV}-cores 4{MiB} shared-{L1} cluster for next-generation open-source software-defined radios,'' in \emph{Proceedings of the Great Lakes Symposium on VLSI 2024}, ser. GLSVLSI '24.\hskip 1em plus 0.5em minus 0.4em\relax New York, NY, USA: ACM, Jun. 2024, p. 86–91.

\bibitem{riedel2023mempool}
S.~Riedel, M.~Cavalcante, R.~Andri, and L.~Benini, ``{MemPool}: A scalable manycore architecture with a low-latency shared {L1} memory,'' \emph{IEEE Transactions on Computers}, vol.~72, no.~12, pp. 3561--3575, 2023.

\bibitem{karlrupp2022}
K.~Rupp, ``50 years of microprocessor trend data,'' \url{https://github.com/karlrupp/microprocessor-trend-data}, 2022.

\bibitem{dinechin2013mppa256}
B.~D. de~Dinechin \emph{et~al.}, ``A clustered manycore processor architecture for embedded and accelerated applications,'' in \emph{2013 IEEE High Performance Extreme Computing Conference (HPEC)}, 2013, pp. 1--6.

\bibitem{dinechin2022}
B.~D. de~Dinechin and L.~Hamon, ``{COOLIDGE}{\texttrademark{}} {MPPA}{\textregistered{}} {DPU} {Kalray’s} unique processor architecture,'' \url{https://www.kalrayinc.com/wp-content/uploads/2023/10/WP_Kalray_MPPA_DPU_Coolidge_june2022.pdf}, Kalray Inc., Tech. Rep., 2022.

\bibitem{sievers2017}
G.~Sievers \emph{et~al.}, \emph{The {CoreVA-MPSoC}: A Multiprocessor Platform for Software-Defined Radio}.\hskip 1em plus 0.5em minus 0.4em\relax Springer, 2017, pp. 29--59.

\bibitem{nvidia2010fermi}
{NVIDIA Corporation}, ``{NVIDIA} {Fermi} {GPU} architecture,'' \url{https://www.nvidia.com/content/pdf/fermi_white_papers/nvidia_fermi_compute_architecture_whitepaper.pdf}, NVIDIA Corp., Tech. Rep., 2010.

\bibitem{nvidia2012kepler}
------, ``{NVIDIA} {GeForce} {GTX} 680,'' \url{https://www.nvidia.com/content/pdf/product-specifications/geforce_gtx_680_whitepaper_final.pdf}, NVIDIA Corp., Tech. Rep., 2012.

\bibitem{nvidia2014maxwell}
------, ``{NVIDIA} {GeForce} {GTX} 750 {Ti},'' \url{https://fabiensanglard.net/cuda/GeForce-GTX-750-Ti-Whitepaper.pdf}, NVIDIA Corp., Tech. Rep., 2014.

\bibitem{nvidia2016pascal}
------, ``{NVIDIA} {Tesla} {P100},'' \url{https://images.nvidia.com/content/pdf/tesla/whitepaper/pascal-architecture-whitepaper.pdf}, NVIDIA Corp., Tech. Rep., 2016.

\bibitem{nvidia2018turing}
------, ``{NVIDIA} {Turing} {GPU} architecture,'' \url{https://images.nvidia.com/aem-dam/en-zz/Solutions/design-visualization/technologies/turing-architecture/NVIDIA-Turing-Architecture-Whitepaper.pdf}, NVIDIA Corp., Tech. Rep., 2018.

\bibitem{nvidia2023ada}
------, ``{NVIDIA} {Ada} {GPU} architecture,'' \url{https://images.nvidia.com/aem-dam/Solutions/Data-Center/l4/nvidia-ada-gpu-architecture-whitepaper-V2.02.pdf}, NVIDIA Corp., Tech. Rep., 2022.

\bibitem{calciu2013}
I.~Calciu \emph{et~al.}, ``Message passing or shared memory: Evaluating the delegation abstraction for multicores,'' in \emph{Principles of Distributed Systems}.\hskip 1em plus 0.5em minus 0.4em\relax Cham: Springer International Publishing, 2013, pp. 83--97.

\bibitem{lang2012}
W.~Lang, S.~Harizopoulos, J.~M. Patel, M.~A. Shah, and D.~Tsirogiannis, ``Towards energy-efficient database cluster design,'' \emph{Proc. VLDB Endow.}, vol.~5, no.~11, pp. 1684--1695, Jul. 2012.

\bibitem{monchiero2006distributedexploration}
M.~Monchiero, G.~Palermo, C.~Silvano, and O.~Villa, ``Exploration of distributed shared memory architectures for noc-based multiprocessors,'' in \emph{2006 International Conference on Embedded Computer Systems: Architectures, Modeling and Simulation}, 2006, pp. 144--151.

\bibitem{sancho2008}
J.~C. Sancho and D.~J. Kerbyson, ``Analysis of double buffering on two different multicore architectures: Quad-core opteron and the cell-be,'' in \emph{2008 IEEE International Symposium on Parallel and Distributed Processing}, 2008, pp. 1--12.

\bibitem{Kalpage2024}
N.~V. Kalpage, P.~Priya, and Y.~Hong, ``{DCT}-based {OTFS} with reduced {PAPR},'' \emph{IEEE Communications Letters}, vol.~28, no.~1, pp. 158--162, 2024.

\bibitem{Li2022}
S.~Li \emph{et~al.}, ``A novel {ISAC} transmission framework based on spatially-spread orthogonal time frequency space modulation,'' \emph{IEEE Journal on Selected Areas in Communications}, vol.~40, no.~6, pp. 1854--1872, 2022.

\bibitem{bertuletti2023pusch}
M.~Bertuletti, Y.~Zhang, A.~Vanelli-Coralli, and L.~Benini, ``Efficient parallelization of {5G-PUSCH} on a scalable {RISC-V} many-core processor,'' in \emph{2023 Design, Automation, and Test in Europe Conference and Exhibition}.\hskip 1em plus 0.5em minus 0.4em\relax Antwerp, Belgium: IEEE, Apr. 2023, pp. 396--401.

\bibitem{bertuletti2023barrier}
M.~Bertuletti, S.~Riedel, Y.~Zhang, A.~Vanelli-Coralli, and L.~Benini, ``Fast shared-memory barrier synchronization for a 1024-cores {RISC-V} many-core cluster,'' in \emph{Embedded Computer Systems: Architectures, Modeling, and Simulation}.\hskip 1em plus 0.5em minus 0.4em\relax Samos: Springer Nature Switzerland, Jul. 2023, pp. 241--254.

\end{thebibliography}
